\documentclass[twocolumn,showpacs,preprintnumbers,amsmath,amssymb]{revtex4}
\usepackage{graphicx}% Include figure files
\usepackage{epsfig}
\usepackage{dcolumn}% Align table columns on decimal point
\usepackage{bm}% bold math
\usepackage{xcolor}% colors
\def\n {\nonumber}

\newcommand{\pp}{{\rm{p-p}}}

\newcommand{\sigmaDPS}{\sigma^{{\rm {DPS}}}}

\newcommand{\sigmaDPSone}{\sigma^{{\rm {DPS,1}}}}
\newcommand{\sigmaDPStwo}{\sigma^{{\rm {DPS,2}}}}
\newcommand{\sigmaDPSthr}{\sigma^{{\rm {DPS,3}}}}
\newcommand{\sigmaSPS}{\sigma^{{\rm {SPS}}}}

\newcommand{\sigmaeffpp}{\sigma_{_{\rm eff,pp}}}

\def\cpc#1#2#3  {{Computer\ Phys.\ Comm.\ }  {\bf#1}, #2 (#3)}
\def\err#1#2#3  {{\it Erratum }              {\bf#1}, #2 (#3)}
\def\epjc#1#2#3 {{Eur. Phys. J. C }          {\bf#1}, #2 (#3)}
\def\dum#1#2#3  {{~}                         {\bf#1}, #2 (#3)}
\def\ib#1#2#3   {{\it ibid. }                {\bf#1}, #2 (#3)}
\def\jcp#1#2#3  {{J.\ Comput.\ Phys.\ }      {\bf#1}, #2 (#3)}
\def\jetpl#1#2#3 {{\rm JETP Lett.}           {\bf#1}, #2 (#3)}
\def\jhep#1#2#3 {{JHEP }                     {\bf#1}, #2 (#3)}
\def\ijmp#1#2#3 {{Int.\ J.\ Mod.\ Phys.\ }   {\bf#1}, #2 (#3)}
\def\jpg#1#2#3  {{J.\ Phys.\ G }             {\bf#1}, #2 (#3)}
\def\mpl#1#2#3  {{Mod.\ Phys.\ Lett.\ }      {\bf#1}, #2 (#3)}
\def\mpla#1#2#3 {{Mod.\ Phys.\ Lett.\ A }    {\bf#1}, #2 (#3)}
\def\ncim#1#2#3 {{Nuovo Cimento }            {\bf#1}, #2 (#3)}
\def\np#1#2#3   {{Nucl.\ Phys.\ }            {\bf#1}, #2 (#3)}
\def\npb#1#2#3  {{Nucl.\ Phys.\ B}           {\bf#1}, #2 (#3)}
\def\pan#1#2#3  {{Phys.\ At.\ Nuclei }       {\bf#1}, #2 (#3)}
\def\plb#1#2#3  {{Phys.\ Lett.\ B }          {\bf#1}, #2 (#3)}
\def\prep#1#2#3 {{Phys.\ Rep.\ }             {\bf#1}, #2 (#3)}
\def\prd#1#2#3  {{Phys.\ Rev.\ D }           {\bf#1}, #2 (#3)}
\def\prl#1#2#3  {{Phys.\ Rev.\ Lett.\ }      {\bf#1}, #2 (#3)}
\def\ptp#1#2#3  {{Prog.\ Theor.\ Phys.\ }    {\bf#1}, #2 (#3)}
\def\ps#1#2#3   {{Physica Scripta }          {\bf#1}, #2 (#3)}
\def\rmp#1#2#3  {{Rev.\ Mod.\ Phys.\ }       {\bf#1}, #2 (#3)}
\def\rpp#1#2#3  {{Rep.\ Prog.\ Phys.\ }      {\bf#1}, #2 (#3)}
\def\sa#1#2#3   {{Sci. Acta}                 {\bf#1}, #2 (#3)}
\def\sjnp#1#2#3 {{Sov.\ J.\ Nucl.\ Phys.\ }  {\bf#1}, #2 (#3)}
\def\spj#1#2#3  {{Sov.\ Phys.\ JETP }        {\bf#1}, #2 (#3)}
\def\spjl#1#2#3 {{Sov.\ JETP Lett.\ }        {\bf#1}, #2 (#3)}
\def\spu#1#2#3  {{Sov.\ Phys.-Usp.\ }        {\bf#1}, #2 (#3)}
\def\yaf#1#2#3  {{Yad.\ Fiz.\ }              {\bf#1}, #2 (#3)}
\def\zp#1#2#3   {{Zeit.\ Phys.\ }            {\bf#1}, #2 (#3)}
\def\zpc#1#2#3  {{Z.\ Phys.\ C }             {\bf#1}, #2 (#3)}

\begin{document}
\title{Double parton scattering versus jet quenching}
\author{S.\ P.\ Baranov}
\email{baranov@sci.lebedev.ru}
\affiliation{P.N. Lebedev Institute of Physics, 
              Lenin Avenue 53, 119991 Moscow, Russia}
\author{A.\ V.\ Lipatov}
\email{lipatov@theory.sinp.msu.ru}
\affiliation{Skobeltsyn Institute of Nuclear Physics,
     Lomonosov Moscow State University, 119991 Moscow, Russia}
\affiliation{Joint Institute for Nuclear Research, Dubna 141980, Moscow
region, Russia}
\author{M.\ A.\ Malyshev}
\email{malyshev@theory.sinp.msu.ru}
\affiliation{Skobeltsyn Institute of Nuclear Physics,
     Lomonosov Moscow State University, 119991 Moscow, Russia}
\affiliation{Joint Institute for Nuclear Research, Dubna 141980, Moscow
region, Russia}     
\author{A.\ M.\ Snigirev}
\email{snigirev@lav01.sinp.msu.ru}
\affiliation{Skobeltsyn Institute of Nuclear Physics,
     Lomonosov Moscow State University, 119991 Moscow, Russia}
\affiliation{Joint Institute for Nuclear Research, Dubna 141980, Moscow
region, Russia}
\date{\today}
\begin{abstract}
A novel observable, the double nuclear modification factor, is proposed to probe simultaneously the initial and final state effects in nucleus-nucleus collisions. 
An interesting competition between the combinatorial enhancement in the double parton scattering and the suppression due to parton energy loss can be observed in the 
production rate of two hard particles. In particular, the production of $J/\psi$ 
mesons in association with a $W$ boson is not suppressed but is enhanced in the 
region of moderate transverse momenta, unlike the case of unassociated (inclusive) 
$J/\psi$ production. 
At the same time, in the region of high enough transverse momenta the 
nuclear modification factor for associated $J/\psi+W$ production converges to that 
of unassociated $J/\psi$.
\end{abstract}
\pacs{12.38.-t, 12.38.Bx}
\maketitle
%----------------------------
\section{Introduction}
%----------------------------

A huge number of intriguing and exquisite observations have been made with the 
Relativistic Heavy Ion Collider (RHIC) and the Large Hadron Collider (LHC). Many
of them could have never been systematically studied at the accelerators of previous
 generation. In particular, hard multiparton interactions (MPI) are just among these
interesting phenomena. The existence of MPI in hadron-hadron collisions at high
energies is a natural consequence of the steep increase in the parton flux at small
parton longitudinal momentum fractions, together with the unitarity requirement for
the cross sections in perturbative QCD. The inclusive cross section of a hard process
is usually calculated under the assumption that, in any collision, along with many
soft interactions, there occurs only a single hard interaction because of its
relatively low probability. Nevertheless, it is also possible that two (or more)
 different parton pairs undergo hard scattering in the same hadronic collision. 
These double parton scatterings (DPS) have been theoretically studied for many 
years, starting from the early days of the parton model. The current state of the 
MPI theory and the results of many years of research have been recently reviewed 
in abook~\cite{Bartalini:2017jkk} (see also a review~\cite{Zinovjev_2021}), which
contains an extensive bibliography. Thus, the investigation of double, triple, and
$n$-parton scatterings~\cite{dEnterria:2016ids,dEnterria:2017yhd} allows us to
extract unique and completely new information about the yet unknown three-dimensional
partonic structure of hadrons and about momentum, flavor, and color correlations in
their wave function.

The DPS events have been initially observed by the AFS (Axial Field Spectrometer)
\cite{AFS} and UA2 (Underground Area 2)~\cite{UA2} collaborations at CERN and later
by the CDF (Collider Detector at Fermilab)~\cite{cdf4jets,cdf} and D0~\cite{D0,D01}
collaborations at the Tevatron, with sufficiently high statistics
for primary analysis and study. As expected, the LHC luminosity and energy provided 
the observation~\cite{Bartalini:2017jkk} of events with hard MPI's in numbers that
are considerably larger than those in the aforementioned experiments. The DPS
contribution has now been reliably separated and measured~\cite{Bartalini:2017jkk} 
in a number of processes containing in the final state jets, gauge bosons ($\gamma,
W, Z$), heavy quarks ($c, b$) and quarkonia ($J/\psi$, $\Upsilon$). For example,
here is the list of some recent results from the collaborations ATLAS (A Toroidal
LHCApparatuS)~\cite{Aaboud:2016dea,Aaboud:2016fzt,Aaboud:2018tiq}, CMS (Compact Muon
Solenoid)~\cite{Chatrchyan:2013xxa,Khachatryan:2015pea,Khachatryan:2016ydm} and LHCb
(Large Hadron Collider beauty)~\cite{Aaij:2012dz,Aaij:2015wpa,Aaij:2016bqq}. 
A triple parton scattering has also been observed very recently~\cite{cms3j-psi}, 
in accordance with an early suggestion~\cite{dEnterria:2016ids,Shao_2019}.

The DPS is actively discussed for proton-nucleus ($p$-$A$) and nucleus-nucleus 
($A$-$A$) collisions as well, since its relative contribution increases, compared
to naive scaling expectation. Unique new options emerge for further studies and
measurements of momentum correlations. The latest achievements in and prospects
for these studies may be found in a review~\cite{dEnterria:2017yhd}. For nucleus-%
nucleus collisions, it opens yet a unique possibility to probe the collective
properties of a new state of dense matter, the quark-gluon plasma (see, e.g.,
\cite{Proceedings:2019drx,qm2019,harris2023qgp}) The experiments at RHIC and the LHC
have provided clear evidence that the production of hadrons in $A$-$A$ collisions
goes through the formation of a fireball of hot and dense quark-gluon plasma. This
follows from the observation of strong suppression of high-$p_T$ particle spectra
(the so-called jet quenching phenomenon expressed in the nuclear modification 
factor $R_{\rm AA}$) and from the results of hydrodynamic simulations of $A$-$A$
collisions.

The main purpose of this Letter is to bring reader's attention to an interesting
possibility to probe the initial and final state effects in nucleus-nucleus
collisions simultaneously, with a single measurement. This can be realised by
introducing a novel observable, the double nuclear modification factor. Our note
is organized as follows. First, we describe our theoretical approach. Then we
show an interesting competition between the combinatorical enhancement in the DPS
and the suppression due to parton energy losses. We illustrate it by the example 
of the associated production of $J/\psi$ mesons and $W$ bosons. 
Some further possible directions of studies are discussed in Conclusions.

%-----------------------------------------
\section{Theoretical setup}

Let us consider nucleus-nucleus collisions at the LHC. 
The parton flux is enhanced by the number $A$ of nucleons in each nucleus, and then
-- modulo (anti)shadowing effects in the nuclear parton distribution functions -- 
the single-parton scattering (SPS) cross section is simply expected to be that of 
\pp\ collisions, or, more exactly, that of nucleon-nucleon collisions ($N$-$N$)
%with N=p,n including protons and neutrons with their appropriate relative fraction) 
scaled by the factor $A^2$, i.e.~\cite{dEnterria:2013mrp}:
\begin{eqnarray} 
\label{eq:sigmaSPSAA}
\!\!\sigmaSPS_{(AA \to a )}\! &\!=\!& \sigmaSPS_{(NN\to a)} 
  \!\int\! {\rm T_A({\bf b_1}) T_A({\bf b_1{-}b})}\,
      d^2{\bf b_1} d^2{\bf b} \nonumber \\[-2mm]
                                   ~&~&~\\[-3mm]
                              &\!=\!& \sigmaSPS_{(NN\to a)} 
  \!\int\! {\rm T_{AA}({\bf b})}\,d^2{\bf b} = A^2 \cdot\sigmaSPS_{(NN\to a)},
  \nonumber
%\label{eq:sigmaSPSAA}
\end{eqnarray}
as long as all of final state effects are out of game.
Here the vector ${\bf b_1}$ measures the distance in the transverse plain
from the center of a nucleus to a given nucleon; 
$\rm{T}_{\rm A}({\bf b_1})$ is the nucleus thickness function which describes the 
nucleon density of a nucleus;
the impact parameter vector ${\bf b}$ connects the centers of the colliding nuclei 
in the transverse plane;
$\rm{T}_{AA}({\bf b})$ is the standard nuclear overlap function normalised to $A^2$,
and $\sigmaSPS_{(NN\to a)}$ is the single inclusive hard scattering cross section.
The normalization constant $A^2$ refers to the combinatorial number of possible 
$N$-$N$ collisions as if the nucleon shadowing effects were neglected.

Traditionally, the level of the energy losses by a hard scattered parton or a
particle $a$ (the final state effects) is quantitatively characterized by the 
nuclear modification factor $R_{\rm AA}(a)$:
\begin{equation} 
R_{\rm AA}(a) =\frac{\sigma_{(AA\to a)}}{A^2\cdot\sigma_{(NN\to a)}}
  \approx \frac{\sigmaSPS_{(AA\to a)}}{A^2\cdot\sigmaSPS_{(NN\to a)}}.
\label{eq:quen1}
\end{equation} 
The left equality gives the definition of the quantity $R_{\rm AA}(a)$;
the right equality is approximate, when the double parton scattering (DPS) processes
are negligible.
The physical meaning of $R_{\rm AA}(a)$ is the survival probability for a particle
produced in a nuclear medium. For a particle untouched by final state interactions 
(such as an electroweak boson, $a=W,\,Z,\,\gamma$) we evidently have $R_{\rm AA}(a)=1$.

We can generalize this nuclear modification factor to the case of two hard scattered 
partons by introducing a double nuclear modification factor:
\begin{equation} 
R_{\rm AA}(a b) =\frac{\sigma_{(AA\to ab)}}{A^2\cdot\sigma_{(NN\to ab)}}
     \approx\frac{\sigmaSPS_{(AA\to ab)}}{A^2\cdot\sigmaSPS_{(NN\to ab)}},
\label{eq:quen2}
\end{equation}
where $\sigmaSPS_{(NN \to a b)}$ and $\sigmaSPS_{(AA \to a b)}$ are the inclusive cross
sections to simultaneously produce two hard particles $a$ and $b$ in an $N$-$N$ or $A$-$A$
collision, respectively.
Similarly to the case of single-particle inclusive production~(\ref{eq:sigmaSPSAA}),
we have a relation for associated $ab$ production:
\begin{eqnarray} 
\!\!\sigmaSPS_{(AA \to ab)}\!\!&\!=\!& \sigmaSPS_{(NN\to ab)} 
  \!\int\! {\rm T_A({\bf b_1}) T_A({\bf b_1{-}b})}\,
      d^2{\bf b_1} d^2{\bf b} \nonumber \\[-2mm]
                                   ~&~&~\\[-3mm]
                             &\!\!=\!& \sigmaSPS_{(NN\to ab)}
  \!\int\!\!{\rm T_{AA}({\bf b})}\,d^2{\bf b} = A^2 \cdot\sigmaSPS_{(NN\to ab)},
  \nonumber
\label{eq:sigmaSPSAA-ab}
\end{eqnarray}
again as long as all of final state effects are out of consideration.

Under a reasonable assumption (approximation) that the produced particles $a$ and $b$
no longer interact with each other and fragment independently, we come to a naive
expectation that the survival probability factorizes:
\begin{equation} 
R_{\rm AA}(ab) = R_{\rm AA}(a)\cdot R_{\rm AA}(b).
\label{eq:quen2-1}
\end{equation}
This may be not fully true in the real life. In the SPS case, the correlations are 
induced by the fact that the both particles are produced in the same point. As so,
they either both have low survival probability if are produced in the inner part of
a nucleus, or both have high survival probability if are produced near the surface.
The particle survival probability may also depend on the kinematic variables.
The particles may either both have large $p_T$ if are produced in a relatively hard collision,
or both have smaller $p_T$ if are produced in a relatively soft collision. 
In view of this, we introduce a numerical prefactor in eq. (\ref{eq:quen2-1})
\begin{equation} 
R_{\rm AA}(ab) = F\cdot R_{\rm AA}(a)\cdot R_{\rm AA}(b)
\label{eq:quen2+}
\end{equation}
with $F={\cal O}(1)$. Recall however that for a particle $a$ untouched by final 
state interactions ($a=W,\,Z,\,\gamma$) we strictly have $R_{\rm AA}(a)=1$ and
$R_{\rm AA}(ab) = R_{\rm AA}(b)$.

The topology of DPS events is more complex and more diverse. Here we have 
a collision of two pairs of partons at a time
\begin{equation}
p_1+p_2\to a,\quad q_1+q_2\to b,
\label{eq:dps}
\end{equation}
and the initial partons can be distributed among the nucleons in various ways.
Accordingly, there are three different contributions showing different dependence
on A.
First, there is a configuration where the partons $p_1$ and $q_1$ are taken from the
same nucleon, and the partons $p_2$ and $q_2$ are taken from another (single) nucleon.
Second is a configuration where the partons $p_1$ and $q_1$ are taken from the same 
nucleon, while the partons $p_2$ and $q_2$ belong to two different nucleons. Third, 
the partons $p_1$, $q_1$, $p_2$, and $q_2$ can all be taken from four different 
nucleons.

The respective contributions to the cross section then read:
\begin{eqnarray} 
\!\!\sigmaDPSone_{(AA\to ab)}\!\!
  &\!=\!&\sigmaDPS_{(NN\to ab)}\!\int\!{\rm T_A({\bf b_1})\,T_A({\bf b_1{-}b})}\,
                                  d^2{\bf b_1}\,d^2{\bf b} \n\\
  &\!=\!&\sigmaDPS_{(NN\to ab)}\!\int\!\!{\rm T_{AA}({\bf b})}\,d^2{\bf b} \n\\
  &\!=\!& A^2 \cdot\sigmaDPS_{(NN\to ab)},
\label{eq:sigmaDPSone}
\end{eqnarray}
\begin{eqnarray} 
\!\!\sigmaDPStwo_{(AA\to ab)}\!\!
  &\!=\!& \frac{m}{2}\cdot\sigmaSPS_{(NN\to a)}\sigmaSPS_{(NN\to b)} \n\\
  &\!\times\!& \,\frac{2(A{-}1)}{A}\int\! {\rm T_A({\bf b_1})\,T_A^2({\bf b_1{-}b})}\,
                                             d^2{\bf b_1}\,d^2{\bf b}\n\\ 
  &\!=\!&\frac{m}{2}\cdot\sigmaSPS_{(NN\to a)}\sigmaSPS_{(NN\to b)}\,2(A{-}1)
               \!\int\! {\rm T_{A}^2({\bf b})}\, d^2{\bf b}    \n\\
  &\!\approx\!& \frac{m}{2}\cdot\sigmaSPS_{(NN\to a)}\sigmaSPS_{(NN\to b)}\,
                                             2 A\,{\rm T_{AA}(0)},
\label{eq:sigmaDPStwo}
\end{eqnarray}
\begin{eqnarray} 
\!\!\sigmaDPSthr_{(AA\to ab)}\!\!
  &\!=\!& \frac{m}{2}\cdot\sigmaSPS_{(NN\to a)}\sigmaSPS_{(NN\to b)} \n\\
  &\!\times\!& \frac{(A{-}1)^2}{A^2} \!\int\! {\rm T_A({\bf b_1})\,T_A({\bf b_1{-}b})}\n\\
  &\!\times\!& {\rm T_A({\bf b_2})\,T_A({\bf b_2{-}b})}\,
                              d^2{\bf b_1}\,d^2{\bf b_2}\,d^2{\bf b} \n\\
  &\!=\!&\frac{m}{2}\cdot\sigmaSPS_{(NN\to a)}\sigmaSPS_{(NN\to b)}\,\frac{(A{-}1)^2}{A^2}
               \!\int\! {\rm T_{AA}^2({\bf b})}\, d^2{\bf b}    \n\\
  &\!\approx\!& \frac{m}{2}\cdot\sigmaSPS_{(NN\to a)}\sigmaSPS_{(NN\to b)}\,A^2\,
                              \frac{{\rm T_{AA}(0)}}{2}
\label{eq:sigmaDPSthr}
\end{eqnarray}
($m=1$ if $a=b$, and $m=2$ if $a$ and $b$ are different).

Now, collecting all the DPS contributions together and using the relation
\begin{eqnarray}
\sigmaDPS_{(NN \to ab)} = \frac{m}{2} \cdot \frac{\sigmaSPS_{(NN \to a )}\cdot
\sigmaSPS_{(NN \to b)}}{\sigmaeffpp}
\label{eq:sps-dps1}
\end{eqnarray}
we get~\cite{dEnterria:2013mrp}:
\begin{equation} 
\sigmaDPS_{(AA\to a b)} = A^2 \,\sigmaDPS_{(NN \to a b)}\cdot C
\end{equation}
with~\cite{dEnterria:2017yhd}
\begin{equation} 
C \approx 1+\frac{\sigmaeffpp \,}{7[\rm mb]\pi}A^{1/3} 
           +\frac{\sigmaeffpp \,}{28[\rm mb]\pi}A^{4/3}, 
\label{eq:doubleAA}
\end{equation}
where the approximation applies to large nuclei.

%{\color{blue}
%The definition of the visible $R_{\rm AA}(ab)$ given by eq. (\ref{eq:quen2}) counts 
%the number of $ab$ pairs produced per $NN$ collision.}

The factor $C$ in Eq.~(\ref{eq:doubleAA}) indicates the enhancement in the DPS cross
sections in $A$-$A$ collisions compared to the corresponding $A^2$-scaled values in
nucleon-nucleon collisions. This enhancement amounts to $C \sim 27$ (for small
$A=40$) or $C \sim 215$ (for large  $A=208$). In all the estimations throughout this
work we use the value of the effective DPS cross section 
$\sigma_\text{eff, pp}=15$~mb.

Notably, the leading terms in Eq.~(\ref{eq:doubleAA}) describe uncorrelated 
production of the particles $a$ and $b$. Indeed, they are produced in different
points and in different partonic subprocesses. 
%Consequently, these contributions satisfy eq. (\ref{eq:quen2-1}).
So, these contributions must basically satisfy eq. (\ref{eq:quen2-1}).

The double nuclear modification factor taking the SPS and DPS contributions 
together can finally be presented as:
\begin{equation}
\label{Rab}
R_{\rm AA}(a b) =R_{\rm AA}(a)\cdot R_{\rm AA}(b)\left[F+\frac{C-F}{K+1}\right],
\end{equation}
where we have introduced a shorthand notation 
\begin{equation}
K=\sigmaSPS_{(NN \to ab)}/\sigmaDPS_{(NN \to ab)}.
\label{eq:sps-dps}
\end{equation}
As an example,
Fig. 1 shows the behavior of $K$-factor for $a=J/\psi$ and $b=W$.

We are going to investigate this novel observable in its
dependence on the type of hard particles ($a$ and $b$) and on the kinematical 
region (mainly the transverse momenta). We can watch an interesting competition
between the combinatorial enhancement $C$ and the $K$- and $R_{\rm AA}$-suppressions. 

%-------------------------------------------
\section{Numerical examples}
%-------------------------------------------

\begin{figure}
\label{fig1}
\epsfig{figure=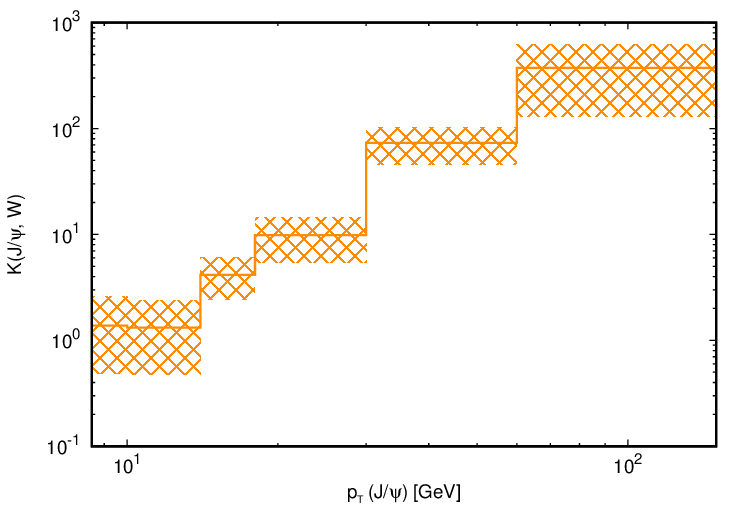, height=50mm, angle=0}
\caption{$K$-factor as a function of $J/\psi$ transverse momentum. The bands show 
uncertainties in the cross section determination for $J/\psi+W$ production.}
\end{figure}

A specific interplay between the effects of DPS and jet quenching can be illustrated
in the simplest case when one of the two hard particles does not lose its energy when
passing through a dense matter. To be more solid, we can employ available
experimental results on the associated production of two hard particles.
Eqs. (\ref{eq:sps-dps1}) and (\ref{eq:sps-dps}) enable us to extract the needed 
$K$-factor directly from the data without appealing to Monte Carlo simulations.
As an example, consider the production of $J/\psi$ mesons in association with a $W$
boson. Relying on the measurements~\cite{Aad_2015} performed by ATLAS collaboration, 
we obtain the $K$-factor as a function of the $J/\psi$ transverse momentum as shown 
in Fig.~1. 
 
Since the $W$ boson passes through the nuclear matter without losing energy, 
our main theoretical prediction~(\ref{Rab}) reduces to
\begin{eqnarray} 
\label{Rjw}
R_{\rm AA}(J/\psi, W) =R_{\rm AA}(J/\psi)\left[1+\frac{C-1}{K(J/\psi, W)+1}\right].
\label{eq:quen2-2}
\end{eqnarray}

The combinatorial enhancement $C$ does not depend on the process kinematics and 
the type of hard particles, but is mainly governed by the 
atomic number $A$. For the minimum bias $Pb$-$Pb$ collisions this enhancement
amounts to $C\sim 215$. The measured nuclear  $J/\psi$ modification factor at the
LHC~\cite{Abelev_2012,Acharya_2020,Sirunyan_2018} amounts to 
$R_{\rm AA}(J/\psi)\simeq 0.5$ at low transverse momenta ($p_T \simeq 2$ GeV/$c$) 
and to $R_{\rm AA}(J/\psi) \simeq 0.3$ over a wide interval of higher transverse
momenta ($p_T> 5$ GeV/$c$). The ``measured'' $K$-factor demonstrates strong
dependence on the transverse momentum: it changes from $\sim 1.4$ at $p_T$ lying 
in the interval [8.5-10] GeV/$c$ to $\sim 374$ at $p_T\in$ [60-150] GeV/$c$. Thus 
we can expect that the production of $J/\psi$ mesons in association with a $W$ 
boson is not suppressed, but is enhanced in the region of moderate transverse
momenta, contrary to unassociated (inclusive) $J/\psi$ production. For $p_T\in$
[8.5-10] GeV/$c$, we have  $$R_{\rm AA}(J/\psi, W) \simeq 30$$
while $R_{\rm AA}(J/\psi) \simeq 0.3$!
In the region of high enough transverse momentum ($p_T > 60$ GeV/$c$), the behavior
of $W$-associated $J/\psi$ production converges to the unassociated case:
$R_{\rm AA}(J/\psi, W) \simeq R_{\rm AA}(J/\psi)$ since the ratio $(C-1)/(K+1)$ 
becomes small. This example clearly demonstrates the competition between the 
effects of DPS (initial state effect) and jet quenching (final state effect). 

The associated production of $D$ mesons and $W$ bosons shows yet a more intriguing
behavior. In this case, there is a notable difference~%
\cite{Baranov:2015nma,Baranov:2016mix} between the opposite-sign and same-sign
production cross sections, and the $K$-factor is significantly larger for $WD$ configurations of the opposite sign than for configurations of the same sign. 
The sensitivity of this factor to the charge configurations  takes place also for other processes with $W$ bosons in the final state~\cite{Kulesza:1999zh,dEnterria:2012jam,Cao:2017bcb}. 
The energy loss is independent of the sign of $D$ mesons: 
$R_{\rm AA}(D^{+})\simeq R_{\rm AA}(D^{-})$.
It means that the double nuclear modification factor for the $WD$ associate
production will be notably larger for the same-sign configurations then for the
opposite-sign ones: 
$$R_{\rm AA}(D^{\pm}, W^{\mp}) < R_{\rm AA}(D^{\pm},W^{\pm})$$ 
at the same kinematics.

\section{Conclusions}
%--------------------------

We propose a novel observable, the double nuclear modification factor, to probe the
initial and final state effects in nucleus-nucleus collisions at a time, ``in one
package''. 
An interesting competition between the combinatorial enhancement due to DPS and the
suppression due to the parton energy losses can be observed. As an illustration we
demonstrate that the production of $J/\psi$ mesons in association with $W$ bosons
is not suppressed but is enhanced in the region of moderate transverse momenta,
unlike the case of unassociated $J/\psi$ production. 
At the same time, in the region of high enough transverse momenta the nuclear
modification factor for associated $J/\psi+W$ production converges to that of
unassociated $J/\psi$.
In the production of $D$ mesons associated with a $W$ boson, the double nuclear
modification factor will be larger for same-sign $WD$ configurations then for
opposite-sign ones at the same kinematics.

We come to the conclusion that measurements of the double nuclear modification
factor potentially open a wide room for further studies of an interplay between 
the effects of DPS and jet quenching, extending to various types of hard
final state particles in a wide interval of their transverse momenta.

The results discussed above assumed ``minimum bias'' $A$-$A$ collisions without any
selection in the reaction centrality. In the future one can also apply this study 
to different centrality classes. The required cross sections for the SPS and DPS
events lying within a certain impact-parameter interval corresponding to a given
centrality percentile may be found in~\cite{dEnterria:2013mrp}.

\acknowledgments
The authors thank A.I.~Demianov, A.V.~Kotikov and I.P.~Lokhtin for useful discussions. A.M.\,Snigirev is grateful to D.~d'Enterria for numerous fruitful discussions during previous work on double parton scattering problems.

\section*{Funding}
The work described in Section~II (Theoretical setup) is supported by the Russian Science Foundation, grant 22-22-00387. The calculation presented in Section~III (Numerical examples) were obtained with the support of the Russian Science Foundation, grant~22-22-00119.

\section*{Conflict of Interest}
The authors of this work declare that they have no conflicts of interest.

\end{document}